\documentclass{aa}
\usepackage[varg]{txfonts}
\usepackage{graphicx}
\usepackage{amsmath,amssymb}
\usepackage[colorlinks=true,citecolor=blue,linkcolor=blue,urlcolor=blue]{hyperref}
\makeatletter\@ifpackageloaded{natbib}{}{\usepackage{natbib}}\makeatother
\newcommand{\Li}{\operatorname{Li}_2}
\newcommand{\Cl}{\operatorname{Cl}_2}
\newcommand{\Clh}{\operatorname{Clh}_2}
\newcommand{\Ti}{\operatorname{Ti}_2}
\providecommand{\arccosh}{\operatorname{arccosh}}
\providecommand{\artanh}{\operatorname{artanh}}
\newcommand{\Cat}{\mathcal{C}}
\newcommand{\los}{{\rm los}}
\newcommand{\ap}{{\rm ap}}
\newcommand{\dd}{\,\mathrm{d}}
\providecommand{\doi}[1]{\href{https://doi.org/#1}{doi:#1}}

\begin{document}

\title{Closed forms for the line-of-sight and aperture velocity dispersions of the isotropic NFW halo}
\titlerunning{Closed-form projected dispersions of the NFW halo}
\author{Artem Andrianov\inst{1}\corrauth{artem.andrianov@evomath.ai}
\and Maarten Baes\inst{2}\email{maarten.baes@ugent.be}}
\authorrunning{A. Andrianov and M. Baes}
\institute{EvoMath Research Labs, Muensterer Str. 49, 65719 Hofheim am Taunus, Germany
\and Department of Physics and Astronomy, Universiteit Gent, Proeftuinstraat 86 N3, B-9000 Ghent, Belgium}
\date{\today}

\abstract
{Analytical dynamical models give physical insight and serve as exact benchmarks for numerical modelling. The
Navarro--Frenk--White (NFW) profile is the most widely used model for dark matter haloes, and many of its properties,
including the radial velocity dispersion and the lensing properties, have long been known in closed form. Its
line-of-sight velocity dispersion $\sigma_\los(R)$ and aperture velocity dispersion $\sigma_\ap(R)$ have not: even for
an isotropic velocity distribution and a tracer that follows the mass, they are evaluated numerically or replaced by fits.}
{We evaluate both projections in closed form.}
{We reduce the projections to single moments, apply Hermite's reduction to their integrands and evaluate the two
remaining integrals with polylogarithm identities. An independent quadrature from the Jeans equation checks the results.}
{Both dispersions are elementary functions of $R/r_s$ plus the same two non-elementary functions, $T_1$ and
$T_2$. On either side of $R=r_s$, $T_1$ is written with Clausen's function or its hyperbolic counterpart, and $T_2$
with the inverse tangent integral. At $R=r_s$,
$\sigma_\los^2=(49/2-3\pi/4-15\,\Cat-12\ln2)\,4\pi G\rho_sr_s^2$, with $\Cat$ Catalan's constant. We give the
small- and large-radius behaviour. The line-of-sight dispersion has its maximum at
$R=0.510049\,r_s$, and the ratio of the aperture dispersion within the virial radius to the virial velocity has its
minimum $0.6154253$ at concentration $c=3.23357$. The independent quadrature agrees with the closed forms to
$2\times10^{-31}$ at 43 radii with $10^{-5}\le R/r_s\le10^4$.}
{The closed forms are exact reference solutions for numerical projections and give exact limits and derivatives. The
combination of classical techniques used here is systematic, and it should extend to stellar tracers in NFW haloes,
whose projected dispersions are what observations measure.}


\maketitle
\nolinenumbers

\section{Introduction}\label{sec:intro}

Analytical models have a long tradition in stellar dynamics, from the model of \citet{plummer1911problem} to the
Jaffe, Hernquist and Dehnen models, the $\eta$-models and the double power-law models of \citet{zhao1996analytical}
\citep{jaffe1983simple,hernquist1990analytical,dehnen1993family,tremaine1994family}, and to more recent families such
as the hypervirial models \citep{evans2005hypervirial} and the models with a power-law logarithmic density slope
\citep{cardone2005spherical}. A model whose potential,
distribution function and velocity moments are known in closed form shows directly how these quantities depend on the
parameters and how they behave at small and large radii
\citep[e.g.][]{baes2002hernquist,baes2005dynamical,baes2007dynamical}. Such models also serve as exact benchmarks for
numerical dynamical modelling. Codes that solve the Jeans equation, project velocity moments or construct equilibrium
models can be tested against them to any precision \citep[e.g.][]{baes2021sphecow}, and so can mass-modelling methods
applied to mock data drawn from known models \citep[e.g.][]{read2021breaking}.

The NFW density profile \citep{navarro1996structure,navarro1997universal}, found in cosmological simulations, is by far
the most widely used analytical model for dark matter haloes. Many of its properties have been known in closed form for
decades: the enclosed mass and the potential, the radial velocity dispersion for an isotropic velocity distribution
(\citealt{lokas2001properties}, hereafter LM01) and for the Mamon--{\L}okas anisotropy profile
\citep{duarte2015maggie}, and the surface density, the projected mass and the lensing properties
\citep{bartelmann1996arcs,wright2000gravitational,golse2002pseudo}. The projected velocity dispersions are an
exception. The line-of-sight dispersion $\sigma_\los(R)$ at projected radius $R$, and the aperture dispersion
$\sigma_\ap(R)$, the surface-density-weighted mean of $\sigma_\los^2$ inside a cylinder of radius $R$, are the
quantities with which dynamical mass estimates of galaxy groups and clusters compare the velocities of the member
galaxies. They follow from the Jeans equation and a projection integral whose origin is usually attributed to
\citet{binney1982ml}. Even for an isotropic velocity distribution and a tracer that follows the mass, they have been written as single
quadratures \citep{tremaine1994family,mamon2005darka,mamon2005darkb}, evaluated numerically (LM01), or replaced by
fitting formulas: \citet{mauduit2007suppressed} (hereafter MM07)
fitted the aperture dispersion at the virial radius with a quartic in $\log c$. Later work uses the same quadratures or
fits \citep{mamon2010universal,verdugo2011gravitational,mamon2013mamposst,sullivan2024predicting}. Thirty years after
the NFW profile was introduced, these projections were among the few basic dynamical properties of the model without
a closed form.

Closed expressions for the projected velocity dispersions are classical for models whose enclosed mass is an algebraic function of radius: the
Hernquist model \citep[his Eqs.~41--42]{hernquist1990analytical}, the $\eta$-models with integer or half-integer $\eta$
\citep{tremaine1994family}, and finite-mass subsets of the double power-law family that contains the NFW profile
\citep[his Eqs.~20--21 and Table~1]{zhao1996analytical}. The NFW halo lies outside these classes: its mass grows like
$\ln(1+r/r_s)$, it has infinite total mass, and its projections lead to the dilogarithm and to Clausen-type functions.

These projections are not hard to evaluate numerically. In units of $4\pi G\rho_sr_s^2$ both depend on $R/r_s$
alone, so one table in $R/r_s$ also serves for fitting, and the NFW profile is itself only an approximation to
simulated haloes. The value of closed forms lies in their exactness: reference values to any precision for testing
numerical projections, exact limits, derivatives in closed form and exact constants. We discuss this, the method and
its extension to other models in Sect.~\ref{sec:limits}.

In this paper we evaluate both quadratures in closed form. Section~\ref{sec:model} recalls the model and
the general expressions for its projected velocity dispersions. Section~\ref{sec:T} gives the two non-elementary
functions on which the results rest, in real forms that mirror each other across $R=r_s$. Section~\ref{sec:J} gives the closed forms of $\sigma_\los$ and
$\sigma_\ap$, their values at $R=r_s$ and their limits. Section~\ref{sec:num} discusses numerical evaluation and
validation, and Sect.~\ref{sec:app} the applications. Section~\ref{sec:limits} discusses what the closed forms add,
the method, and its extension to tracer populations in NFW haloes and to other models. The derivations are in the
appendices. The closed forms were derived by hand and with computer algebra
(SymPy), and checked with arbitrary-precision numerics (mpmath). Given the functions $T_1$ and
$T_2$ of Sect.~\ref{sec:T} as input columns, the EvoMath symbolic-regression system \citep{evomath} later re-found the transcendental part
of the line-of-sight integral from numerical tables; it did not recover the full integral or its aperture
counterpart. The derivations do not rely on this search.

\section{The NFW model and its projected velocity dispersions}\label{sec:model}

\subsection{The model}\label{sec:nfw}

We consider a spherical halo with the NFW density profile, a tracer population that follows the mass, and an isotropic
velocity distribution. Throughout the paper we use units in which $r_s=\rho_s=G=1$, where $r_s$ is the scale radius and
$\rho_s$ the characteristic density of the halo. Radii are then in units of $r_s$, masses in units of $\rho_sr_s^3$,
and squared velocities in units of $G\rho_sr_s^2$. In these units the density and the enclosed mass are
\begin{equation}\label{eq:rhoM}
\rho(r)=\frac{1}{r\,(1+r)^2},\quad M(r)=4\pi\Big[\ln(1+r)-\frac{r}{1+r}\Big].
\end{equation}
The surface density at projected radius $R$ and the projected mass within $R$ are elementary
(\citealt{bartelmann1996arcs}; LM01 Eqs.~41--43):
\begin{align}
\Sigma(R)&=2\int_R^\infty\frac{\rho(r)\,r\dd r}{\sqrt{r^2-R^2}}=\frac{2\,[1-F(R)]}{R^2-1},\label{eq:Sigma}\\
M_p(R)&=2\pi\int_0^R\Sigma(R')\,R'\dd R'=4\pi\Big[\ln\frac R2+F(R)\Big],\label{eq:Mp}
\end{align}
with
\begin{equation}\label{eq:F}
F(R)=\begin{cases}\dfrac{\arccosh(1/R)}{\sqrt{1-R^2}}, & R<1,\\[2ex]
\dfrac{\arccos(1/R)}{\sqrt{R^2-1}}, & R>1,\end{cases}
\end{equation}
and $F(1)=1$. All three functions are smooth at $R=1$, where $\Sigma(1)=\frac23$ and $M_p(1)=4\pi(1-\ln2)$
(Appendix~\ref{app:Ffh}).

For an isotropic velocity distribution the Jeans equation gives the radial velocity dispersion
\begin{equation}\label{eq:jeans}
\sigma_r^2(r)=\frac{1}{\rho(r)}\int_r^\infty\frac{\rho(r')\,M(r')}{r'^2}\dd r' ,
\end{equation}
which LM01 (Eq.~14) evaluated in closed form with the dilogarithm; we do not need the explicit expression.

The halo is not truncated: the Jeans equation is integrated with zero pressure at infinity, the projections run over
the whole line of sight, and the aperture is a cylinder of radius $R$ with unbounded length. A halo truncated at a
finite radius changes the pressure, the surface density and the projected mass, and is outside the scope of this paper.

\subsection{Line-of-sight and aperture velocity dispersions}\label{sec:proj}

The line-of-sight velocity dispersion at projected radius $R$, and the aperture velocity dispersion, which is the
surface-density-weighted mean of $\sigma_\los^2$ within $R$, are
\begin{align}
\sigma_\los^2(R)&=\frac{2}{\Sigma(R)}\int_R^\infty\frac{\rho(r)\,\sigma_r^2(r)\,r\dd r}{\sqrt{r^2-R^2}},\label{eq:defs1}\\
\sigma_\ap^2(R)&=\frac{2\pi}{M_p(R)}\int_0^R\Sigma(R')\,\sigma_\los^2(R')\,R'\dd R'\label{eq:defs2}
\end{align}
(LM01 Eq.~44 at $\beta=0$, and Eqs.~46--47). Inserting \eqref{eq:jeans} and exchanging the order of integration
reduces both to single integrals:
\begin{equation}\label{eq:obs}
\sigma_\los^2(R)=\frac{8\pi\,J_1(R)}{\Sigma(R)},\quad
\sigma_\ap^2(R)=\frac{16\pi^2}{3}\,\frac{\tfrac12-J_3(R)}{M_p(R)},
\end{equation}
where, for $k=1$ and $k=3$,
\begin{equation}\label{eq:Jk}
\begin{split}
J_k(R)&=\frac{1}{4\pi}\int_R^\infty\frac{\rho(r)\,M(r)}{r^2}\,\big(r^2-R^2\big)^{k/2}\dd r\\
&=\int_R^\infty\Big[\ln(1+r)-\frac{r}{1+r}\Big]\frac{(r^2-R^2)^{k/2}}{r^3(1+r)^2}\dd r .
\end{split}
\end{equation}
Indeed, the integrands are positive, so that
\begin{multline*}
\int_R^\infty\frac{r\dd r}{\sqrt{r^2-R^2}}\int_r^\infty\frac{\rho(r')\,M(r')}{r'^2}\dd r'\\
=\int_R^\infty\frac{\rho(r')\,M(r')}{r'^2}\sqrt{r'^2-R^2}\dd r'=4\pi J_1(R),
\end{multline*}
which gives the first relation. For the second, $J_3'(R)=-3R\,J_1(R)$ and
$J_3(0)=(4\pi)^{-1}\int_0^\infty\rho\,M\,r\dd r=\frac12$, so that
$\int_0^RJ_1(R')\,R'\dd R'=\frac13[\frac12-J_3(R)]$.

Both relations hold for any isotropic spherical model, with $\rho\,M$ replaced by the product of the tracer density
and the total mass: they are the third line of Eq.~29 and the numerator of Eq.~37 of \citet{tremaine1994family}. The
first is also the isotropic case of Eq.~A15 of \citet{mamon2005darkb}, whose kernel covers the usual anisotropy
profiles, and the second is Eq.~B7 of \citet{mamon2005darka} and Eq.~A.2 of MM07, in both cases with the factor $1/3$
restored by the errata \citep{mamon2006erratum,mauduit2009erratum}. The denominator of Eq.~37 of
\citet{tremaine1994family} is a difference of two integrals that both diverge for the NFW halo, so we use the
projected mass \eqref{eq:Mp} instead.

In physical units $R$ stands for $R/r_s$, and the right-hand sides of \eqref{eq:obs} are multiplied by
$G\rho_sr_s^2$, so that both squared dispersions scale with $4\pi G\rho_sr_s^2$. What remains is to evaluate $J_1$
and $J_3$.

\section{The functions $T_1$ and $T_2$}\label{sec:T}

The projected dispersions can be expressed in terms of elementary functions and two non-elementary functions,
\begin{align}
T_1(R)&=\int_R^\infty\frac{\ln(1+r)}{1+r}\,\frac{\dd r}{\sqrt{r^2-R^2}},\notag\\
T_2(R)&=\int_R^\infty\frac{\ln(1+r)}{r}\,\frac{\dd r}{\sqrt{r^2-R^2}}.\label{eq:Tdef}
\end{align}
Both are projection integrals of the same kind as the surface density \eqref{eq:Sigma}. They can be written with three
classical functions: Clausen's function, its hyperbolic counterpart, and the inverse tangent integral,
\begin{align}
\Cl(z)&=-\int_0^z\ln\Big(2\sin\frac t2\Big)\dd t&&(0\le z\le2\pi),\label{eq:Cl}\\
\Clh(z)&=-\int_0^z\ln\Big(2\sinh\frac t2\Big)\dd t&&(z\ge0),\label{eq:Clh}\\
\Ti(z)&=\int_0^z\frac{\arctan t}{t}\dd t&&(z\ \mbox{real})\label{eq:Ti}
\end{align}
(\citealt{lewin1981polylogarithms}; \citealt{dlmf}, Sect.~25.12; for the hyperbolic Clausen function see
\citealt{binger2006form}, their Appendix~C). All three are relatives of the dilogarithm
$\Li(z)=-\int_0^zt^{-1}\ln(1-t)\dd t$:
\begin{gather}
\Cl(z)=\operatorname{Im}\Li\big(e^{iz}\big),\qquad\Ti(z)=\operatorname{Im}\Li(iz),\notag\\
\Clh(z)=\frac{\pi^2}{6}-\frac{z^2}{4}-\Li\big(e^{-z}\big)\label{eq:rel}
\end{gather}
(Appendix~\ref{app:T}). At the special arguments that occur below, $\Ti(1)=\Cl(\pi/2)=\Cat$, where
$\Cat=0.9159655941\ldots$ is Catalan's constant.

Let $\vartheta$ denote the numerator of $F$ in \eqref{eq:F}, that is $\vartheta=\arccosh(1/R)$ for $R\le1$ and
$\vartheta=\arccos(1/R)$ for $R\ge1$, and let $\chi=\arccosh R$ for $R\ge1$. Then
\begin{equation}\label{eq:T1}
\begin{split}
T_1(R)={}&F(R)\,\ln\frac{2\,|1-R^2|}{R}\\
&+\frac{1}{\sqrt{|1-R^2|}}\begin{cases}
\Clh(2\vartheta), & R<1,\\[0.8ex]
\Cl(2\vartheta), & R>1,
\end{cases}
\end{split}
\end{equation}
\begin{equation}\label{eq:T2}
T_2(R)=\frac1R\begin{cases}
2\Ti\big(e^{-\vartheta}\big)+\vartheta\arcsin R, & R\le1,\\[1ex]
2\Ti\big(e^{-\chi}\big)+\dfrac{\pi\chi}{2}, & R\ge1,
\end{cases}
\end{equation}
with $T_1(1)=2-\ln2$ and $T_2(1)=2\Cat$. The arguments of the inverse tangent integral are
$e^{-\vartheta}=R/(1+\sqrt{1-R^2})$ and $e^{-\chi}=1/(R+\sqrt{R^2-1})$. The derivations are in
Appendix~\ref{app:T}.

The two branches of $T_1$ mirror each other in the same way as those of $F$, which appears in \eqref{eq:T1}
because $\vartheta=\sqrt{|1-R^2|}\,F(R)$: across $R=1$ the hyperbolic functions turn into the trigonometric ones. Both $T_1$ and $T_2$ extend to holomorphic functions on $\operatorname{Re}R>0$, so
each pair of branches belongs to one real-analytic function, and the formulas are continuous through $R=1$. The
function $T_2$ is tied to the surface density through $F$: $\frac{\dd}{\dd R}[R\,T_2(R)]=F(R)$, so that
$R\,T_2(R)=\int_0^RF(R')\dd R'$. Other forms are possible; for $R<1$, for instance,
$\sqrt{1-R^2}\,T_1(R)=\Li(1-a^2)+\ln a\,\ln(1+a^2)$ with $a=e^{-\vartheta}$ (Appendix~\ref{app:Tcont}).

\section{Closed forms of the projected dispersions}\label{sec:J}

A Hermite reduction of the integrands of \eqref{eq:Jk} and an integration by parts (Appendix~\ref{app:J}) give, for every
$R>0$,
\begin{equation}\label{eq:J1}
\begin{split}
J_1(R)={}&(3R^2-2)\,T_1(R)-\Big(3R^2-\frac12\Big)T_2(R)\\
&+\frac{11+R^2}{2}-\frac{\pi R}{4}+3\ln\frac R2\\
&+\frac{5-3R^2-R^4}{4}\,\Sigma(R),
\end{split}
\end{equation}
\begin{equation}\label{eq:J3}
\begin{split}
J_3(R)={}&3R^2(1-R^2)\,T_1(R)+\frac32R^2(2R^2-1)\,T_2(R)\\
&+\frac{3-11R^2-R^4}{2}+\frac{\pi R^3}{4}+(1-3R^2)\ln\frac R2\\
&+\frac{R^6+3R^4-6R^2+2}{4}\,\Sigma(R) .
\end{split}
\end{equation}
All coefficients are polynomials in $R$, and all terms are real-analytic on $R>0$, so the formulas also hold at $R=1$.
Inserted in \eqref{eq:obs}, with $\Sigma(R)$ and $M_p(R)$ from \eqref{eq:Sigma} and \eqref{eq:Mp}, they give the
line-of-sight and aperture velocity dispersions of the NFW halo in closed form.

Since $J_3'=-3R\,J_1$ and $M_p'=2\pi R\,\Sigma$, the derivative of $\sigma_\ap^2$ with respect to $R$ is again in
closed form. The derivatives $T_1'$ and $T_2'$ are elementary combinations of $T_1$, $T_2$ and logarithms, so the
derivative of $\sigma_\los^2$ and the circular-orbit dispersion (LM01 Eq.~45) can be expressed in the same functions;
we do not write them out.

\subsection{Values at the scale radius and limits}\label{sec:values}

At the scale radius, $T_1(1)=2-\ln2$, $T_2(1)=2\Cat$, $\Sigma(1)=\frac23$ and $M_p(1)=4\pi(1-\ln2)$ give
\begin{align}
\sigma_\los^2(1)&=4\pi\Big(\frac{49}{2}-\frac{3\pi}{4}-15\,\Cat-12\ln2\Big),\label{eq:rs}\\
\sigma_\ap^2(1)&=4\pi\,\frac{5-\frac\pi4-3\,\Cat-2\ln2}{3\,(1-\ln2)},\notag
\end{align}
that is $\sigma_\los^2(1)=0.0865554304300\times4\pi$ and $\sigma_\ap^2(1)=0.0873499039895\times4\pi$. The two
dispersions are close there: $\sigma_\los=0.29420$ and $\sigma_\ap=0.29555$ in units of
$(4\pi G\rho_sr_s^2)^{1/2}$.

At small radii (Appendix~\ref{app:limits}),
\begin{align}
\sigma_\los^2(R)&\simeq4\pi\,\frac{\frac72-\frac{\pi^2}{3}-\frac{\pi R}{4}}{\ln(2/R)-1},\notag\\
\sigma_\ap^2(R)&\simeq4\pi\,\frac{\frac72-\frac{\pi^2}{3}-\frac{\pi R}{6}}{\ln(2/R)-\frac12},\label{eq:small}
\end{align}
with relative errors of order $R^2\ln(1/R)$, which amount to $0.2\%$ and $0.1\%$ at $R=0.01$. Both dispersions
therefore tend to zero at the centre, but only like $1/\ln(1/R)$. We keep the complete denominators and the terms
linear in $R$ because the leading term alone, $4\pi(\frac72-\frac{\pi^2}{3})/\ln(1/R)$, has relative corrections of
order $1/\ln(1/R)$: it is still in error by $3\%$ and $2\%$ at $R=10^{-4}$. At large radii,
\begin{equation}\label{eq:large}
\sigma_\los^2(R)\simeq\frac{\pi^2}{4R}\Big[\ln(2R)-\frac54\Big],\quad
\sigma_\ap^2(R)\simeq\frac{2\pi}{3\ln(R/2)},
\end{equation}
with relative errors of order $1/R$ and $\ln R/R$. Appendix~\ref{app:limits} gives the corresponding expansions of
$J_1$ and $J_3$, their values at $R=1$, and the next term of $J_1$ at small radii.

The law $\sigma^2\propto1/\ln(1/R)$ at small radii is not specific to the NFW halo. \citet{zhao1996analytical} gives it
for the finite-mass members of his family with an inner slope of $1$ (his Eq.~35), and for the Hernquist model it
follows from Eq.~41 of \citet{hernquist1990analytical} (the leading term of his Eq.~B7 is corrected in footnote~3 of
\citealt{tremaine1994family}). For the NFW halo, \eqref{eq:small} also gives the constant $\frac72-\frac{\pi^2}{3}$
and the terms linear in $R$.

The limits have a direct physical reading. At small $R$ the projected pressure $\Sigma\sigma_\los^2=8\pi J_1$ tends
to the finite value $8\pi(\frac72-\frac{\pi^2}{3})$, that is $8\pi G\rho_s^2r_s^3(\frac72-\frac{\pi^2}{3})$ in
physical units. The three-dimensional pressure $\rho\sigma_r^2$ diverges at the centre, but only like
$2\pi\ln(1/r)$, and is integrable along the line of sight, while the surface density diverges logarithmically. At
large $R$ the aperture dispersion decreases only logarithmically, because the untruncated aperture contains the whole
halo along the line of sight, whose mass grows logarithmically. For a truncated halo this behaviour is different.

\section{Numerical evaluation and validation}\label{sec:num}

\paragraph{Double precision.} Evaluated in double precision (NumPy, with $\Li$
and $\Cl$ from the complex \texttt{spence} function of SciPy) and compared with a 30-digit reference, the closed form
for $J_1$ loses digits in two places. Near $R=1$, $\Sigma=2(1-F)/(R^2-1)$ is a quotient of small numbers. With
$\Sigma$ computed as written, the relative error of $J_1$ is $4.5\times10^{-10}$ at $R=0.999$ and $2.6\times10^{-11}$
at $1.001$; with $\Sigma$ from its series (Appendix~\ref{app:Ffh}) for $|1-R^2|<0.05$ it is $1.3\times10^{-12}$ and
$2.4\times10^{-14}$. At large $R$, terms of order $R^3$ cancel to a result of order $\ln R/R^3$, and the error grows to
$8.9\times10^{-11}$, $8.0\times10^{-10}$ and $1.9\times10^{-6}$ at $R=10$, $30$ and $100$. Elsewhere it is between
$8\times10^{-15}$ and $7\times10^{-13}$ for $0.01\le R\le3$. A double-precision implementation therefore needs the
series for $\Sigma$ near $R=1$ and, for $R\gtrsim30$, the large-radius expansions of Appendix~\ref{app:limits} or
extended precision. \citet[App.~B]{hernquist1990analytical} met the same problem at $R=a$ in his elementary projection
and used Taylor expansions there. Table~\ref{tab:values} lists reference values.

\begin{table*}
\caption{Values of the closed forms, computed with 120 working digits and rounded to 12. In physical units $R$ is in
units of $r_s$, and the last two columns are the squared dispersions in units of $4\pi G\rho_sr_s^2$.}
\label{tab:values}
\centering
\begin{tabular}{rllll}
\hline\hline
$R$ & $J_1$ & $J_3$ & $\sigma_\los^2/4\pi$ & $\sigma_\ap^2/4\pi$\\
\hline
$0.01$ & $0.202684544890$ & $0.499969233557$ & $0.0471470454589$ & $0.0427430050838$\\
$0.1$ & $0.153245413650$ & $0.497455779275$ & $0.0755428955373$ & $0.0674687287935$\\
$0.5$ & $0.0627119393828$ & $0.465590045043$ & $0.0903297058633$ & $0.0853436541947$\\
$1$ & $0.0288518101433$ & $0.419589307049$ & $0.0865554304300$ & $0.0873499039895$\\
$2$ & $0.00990786875522$ & $0.347251220583$ & $0.0751734707505$ & $0.0842148158330$\\
$10$ & $3.25155682681\times10^{-4}$ & $0.157204825426$ & $0.0377734743596$ & $0.0650252378932$\\
$100$ & $7.87090919661\times10^{-7}$ & $0.0295999276836$ & $0.00799491263478$ & $0.0399222824813$\\
\hline
\end{tabular}
\end{table*}

\begin{figure}
\centering
\includegraphics[width=\hsize]{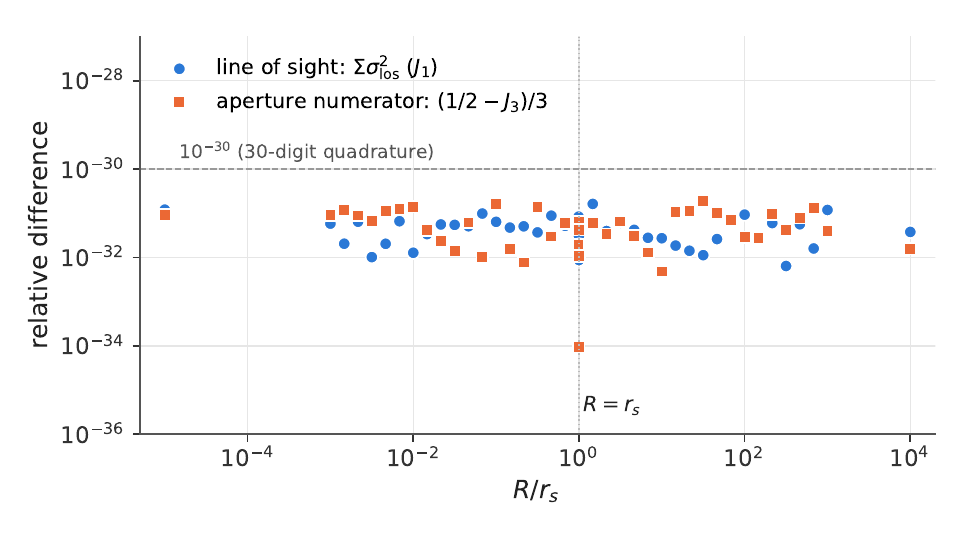}
\caption{Relative difference between the closed forms and the definition-based route (radial pressure from the Jeans
equation by quadrature, then line-of-sight or cylinder integration), at 30 working digits. Circles:
$\Sigma\sigma_\los^2$, that is $J_1$. Squares: the aperture numerator $\frac13(\frac12-J_3)$. The dotted line marks
the scale radius.}
\label{fig:validation}
\end{figure}

\paragraph{Validation.} The derivations do not use numerics. The following checks test the printed formulas for
algebra and branch errors. The substitution identities, the antiderivative of Appendix~\ref{app:T1low} and the
rational identities of Table~\ref{tab:hermite} are verified exactly with SymPy. A definition-based route computes the
radial pressure $\rho\sigma_r^2$ from the Jeans equation \eqref{eq:jeans} by quadrature, then $\Sigma\sigma_\los^2$ by
integrating it along the line of sight, and the aperture numerator by integrating it over the volume of the cylinder.
It also computes $\Sigma$ and $M_p$ from their definitions as integrals of $\rho$. None of the steps of
Sects.~\ref{sec:T} and~\ref{sec:J}, nor the exchange of integration order behind \eqref{eq:obs}, enters this route.
With mpmath tanh--sinh quadrature at 30 working digits, it agrees with the closed forms to $1.6\times10^{-31}$ for
$J_1$, $1.9\times10^{-31}$ for the aperture numerator and $1.2\times10^{-31}$ for $\Sigma$ and $M_p$, at 43 radii from
$R=10^{-5}$ to $10^4$ that include $R=1$ and $|R-1|=10^{-3}$ and $10^{-2}$ (Fig.~\ref{fig:validation}). A 50-digit
quadrature of \eqref{eq:Tdef} agrees with \eqref{eq:T1} and \eqref{eq:T2} to $2\times10^{-49}$ at 14 radii from
$R=10^{-3}$ to $10^3$. The scripts are described in Appendix~\ref{app:scripts}.

\section{Applications}\label{sec:app}

\begin{figure}
\centering
\includegraphics[width=\hsize]{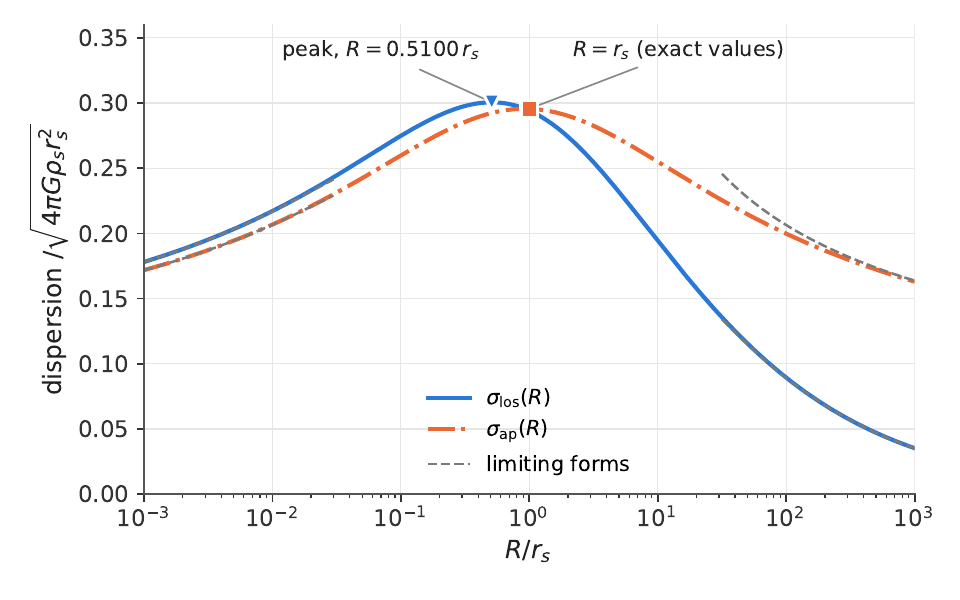}
\caption{Line-of-sight dispersion $\sigma_\los(R)$ (solid) and aperture dispersion $\sigma_\ap(R)$ (dash-dotted) of
the isotropic, untruncated NFW halo, in units of $(4\pi G\rho_sr_s^2)^{1/2}$. Markers: the exact values at the scale
radius \eqref{eq:rs}, $0.29420$ and $0.29555$, and the maximum of $\sigma_\los$ at $R=0.510049\,r_s$. Dashed: the
limiting forms \eqref{eq:small} for $R\le0.03\,r_s$ and \eqref{eq:large} for $R\ge30\,r_s$ (square roots plotted).}
\label{fig:profiles}
\end{figure}

All numbers in this section are values of the closed forms at points located numerically, computed with 30 to 120
significant digits. Figure~\ref{fig:profiles} shows both dispersions for $10^{-3}\le R\le10^3$.

\paragraph{Maximum of $\sigma_\los$.} On $10^{-3}\le R\le10^3$ the line-of-sight dispersion has a single maximum, at
$R_{\rm pk}=0.510048783501$, where $\sigma_\los^2=0.0903328392478\times4\pi$.
In units of $r_s$ this position does not depend on the concentration.

\paragraph{Aperture-to-virial ratio.} For a halo of concentration $c=r_{\rm vir}/r_s$ the virial radius is $r=c$ in
our units, and the virial velocity is given by $V_{\rm vir}^2=M(c)/c$. The ratio of the aperture dispersion inside the
cylinder of radius $r_{\rm vir}$ to the virial velocity is then
\begin{equation}\label{eq:eta}
\eta^2(c)\equiv\frac{\sigma_\ap^2(c)}{V_{\rm vir}^2}=\frac{16\pi^2}{3}\,\frac{c\,\big[\tfrac12-J_3(c)\big]}{M_p(c)\,M(c)} .
\end{equation}
This is the quantity of Eq.~A.9 of MM07 at $Y=c$ for the isotropic, untruncated halo. Since $J_3'=-3R\,J_1$,
$M_p'=2\pi R\,\Sigma$ and $M'=4\pi r^2\rho$, the derivative of $\eta^2$ is in closed form, and its root gives the
minimum
\[
c_*=3.23357476264369,\qquad \eta(c_*)=0.61542532466637 .
\]
MM07 state that the ratio reaches a minimum of $0.62$ for $2<c<5$, consistent with these digits. In terms of the
virial quantities, $R/r_s=cR/r_{\rm vir}$ and $G\rho_sr_s^2=c\,V_{\rm vir}^2/M(c)$, so Fig.~\ref{fig:profiles} also
gives $\sigma_\los/V_{\rm vir}$ and $\sigma_\ap/V_{\rm vir}$ after a shift in $\log R$ and a rescaling.

\paragraph{Accuracy of the Mauduit--Mamon fit.} MM07 (their Eq.~A.10) approximate $\eta(c)$ by
$\log_{10}\eta=-0.1539-0.2138L+0.2358L^2-0.05357L^3+0.005515L^4$ with $L=\log_{10}c$, and state an accuracy better
than $5\%$ for $c>0.5$ and better than $0.12\%$ for $c>1$; the erratum \citep{mauduit2009erratum} leaves this fit
unchanged. Table~\ref{tab:mm07} compares the fit with the exact $\eta(c)$. For $c>1$ the fit
deviates by $0.35$ to $1.4\%$ for $2\le c\le4$ and by $2.3\%$ at $c=1.5$. The deviation falls below $0.12\%$ only for
$c>5.68$. It stays below $0.09\%$ for $6.1\le c\le30$, but exceeds $0.12\%$ again for $41.5<c<84.5$ (up to $0.14\%$ at
$c\approx60$), for $251<c<687$ and for $c>955$. At $c=1$ the deviation is $4.3\%$, and for $0.5<c<0.907$ it exceeds
$5\%$ (up to $11.5\%$ at $c=0.5$). MM07 also quote $\sigma_1=433$, $436$ and $463\ {\rm km\,s^{-1}}$ for $c=4$, $5$ and
$10$, which in their normalisation equal $700\,\eta\ {\rm km\,s^{-1}}$; the exact ratio gives $432$, $436$ and $463$.
For the isotropic case, Eq.~\eqref{eq:eta} can be used in place of the fit. For the Mamon--{\L}okas anisotropy used in MAMPOSSt
\citep{mamon2013mamposst} the corresponding ratio is not covered here.

\begin{table*}
\caption{Exact aperture-to-virial ratio $\eta(c)$, Eq.~\eqref{eq:eta}, and the fit of MM07 (their Eq.~A.10).}
\label{tab:mm07}
\centering
\begin{tabular}{rlllcrlll}
\hline\hline
$c$ & $\eta$ (exact) & MM07 fit & fit/exact $-1$ & & $c$ & $\eta$ (exact) & MM07 fit & fit/exact $-1$\\
\hline
$0.5$ & $0.7691433869$ & $0.85769882$ & $+0.1151$ & & $5$ & $0.6224015649$ & $0.62357012$ & $+0.0019$\\
$1$ & $0.6724918563$ & $0.70161683$ & $+0.0433$ & & $6$ & $0.6293277374$ & $0.62993162$ & $+0.00096$\\
$1.5$ & $0.6390303020$ & $0.65384710$ & $+0.0232$ & & $7$ & $0.6369736711$ & $0.63724210$ & $+0.00042$\\
$2$ & $0.6244457457$ & $0.63341991$ & $+0.0144$ & & $10$ & $0.6608793140$ & $0.66076191$ & $-0.00018$\\
$3$ & $0.6156384552$ & $0.61977982$ & $+0.0067$ & & $20$ & $0.7324897510$ & $0.73265854$ & $+0.00023$\\
$4$ & $0.6171085440$ & $0.61927760$ & $+0.0035$ & & $30$ & $0.7907332200$ & $0.79137799$ & $+0.00082$\\
\hline
\end{tabular}
\end{table*}

\section{Discussion}\label{sec:limits}

\subsection{Scope}\label{sec:scope}

The results hold for the isotropic, spherical, untruncated NFW halo with a tracer that follows the mass; a different
tracer profile, anisotropy or truncation changes the integrals. To our knowledge, a closed form of the isotropic NFW
line-of-sight or aperture dispersion has not been given before. With it, the projected velocity dispersions join the
surface density, the projected mass and the lensing properties among the properties of the model that are known in
closed form.

\subsection{What the closed forms add}\label{sec:value}

Line-of-sight projections and aperture integrations are routine steps in dynamical modelling, in Jeans codes as well
as in the projection of numerical equilibrium models, and the NFW halo is among the most common test cases. Their
numerical evaluation has known pitfalls: the integrable singularity of the projection kernel at $r=R$, the truncation
of the line-of-sight integral at a finite radius, and the slow convergence of the aperture integral of a halo of
infinite mass. Errors of this kind are hard to detect against another numerical result. Equations~\eqref{eq:J1},
\eqref{eq:J3} and~\eqref{eq:obs}, or Table~\ref{tab:values}, give exact reference values instead, to any precision.

The closed forms also give exact information that a table does not. The central value $8\pi(\frac72-\frac{\pi^2}{3})$
of the projected pressure $\Sigma\sigma_\los^2$, the $1/\ln(1/R)$ approach of both dispersions to zero, the large-radius forms
\eqref{eq:large}, the value at $R=r_s$ with Catalan's constant, the position of the maximum of $\sigma_\los$ and the
minimum of $\eta(c)$ follow from them directly. Derivatives with respect to the projected radius and to the
concentration are again in closed form (Sects.~\ref{sec:J} and~\ref{sec:app}), which helps in fits that use gradients
and in Fisher-matrix forecasts.

The comparison with the fit of MM07 in Sect.~\ref{sec:app} is an example of the first use. It does not need the closed
form as such: a careful quadrature at each concentration shows the same deviations. It illustrates, however, that
fitting formulas calibrated on numerical integrations are worth checking against exact values. We also note what the
closed forms do not give: in double precision they are not more accurate than a good adaptive quadrature, and near
$R=r_s$ and at large radii they need the series and expansions of Sect.~\ref{sec:num}.

\subsection{The method}\label{sec:method}

The results do not come from recognising an integral in a table. They follow from four classical steps, each of which
can be checked separately:
\begin{enumerate}
\item The projection identities of Sect.~\ref{sec:model} \citep{tremaine1994family,mamon2005darka,mamon2005darkb}
reduce the line-of-sight and aperture integrals of the Jeans pressure to the single moments $J_1$ and $J_3$ of $\rho M$
against odd powers of the projection kernel $\sqrt{r^2-R^2}$.
\item Hermite's reduction \citep{hermite1872integration,bronstein2005symbolic} splits the algebraic part of each
integrand, a rational function times $(r^2-R^2)^{k/2}$, into an exact derivative and two simple terms, proportional
to $1/[(1+r)\sqrt{r^2-R^2}]$ and $1/[r\sqrt{r^2-R^2}]$ (Appendix~\ref{app:J}). Multiplied by $\ln(1+r)$, these are
the integrands of $T_1$ and $T_2$, for both moments.
\item Substitutions that rationalise the kernel, $r=\frac R2(u+u^{-1})$, $r=R\sec\psi$ and $r=R\cosh t$, turn $T_1$
and $T_2$ into integrals of logarithms of rational and trigonometric functions (Appendix~\ref{app:T}).
\item The functional equations of the dilogarithm \citep{lewin1981polylogarithms,dlmf}, together with analytic
continuation across $R=1$, reduce these integrals to $\Cl$, $\Clh$ and $\Ti$ in real forms on either side of the
scale radius.
\end{enumerate}
Step~2 is the reason why both projected moments, and the derivatives of both dispersions, are expressed in the same two
non-elementary functions, with polynomial coefficients. The route also predicts which functions can appear. The
integrand contains one logarithm, from $M(r)$, and the integration adds one more, so the result has weight two in the
sense of polylogarithms: dilogarithms and their relatives, but no trilogarithms.

None of these techniques is new, and we do not claim a new integration method;
\citet[Sect.~4.2]{mancino2019structural}, for instance, used Hermite's reduction for the radial dispersions of the J3
models. Their combination, however, gives a systematic route from the projection integrals of a model with a
logarithmic mass profile to a small basis of special functions. In a similar spirit, \citet{baes2011sersic} and
\citet{baes2011sersicII} used Mellin integral transforms to write the deprojected S\'ersic model in terms of the Meijer
G and Fox H functions, in part to show the usefulness of a technique that had rarely been used in the field. The route
used here may be useful in the same way for other projections in stellar dynamics.

\subsection{Tracers in NFW haloes and other models}\label{sec:other}

The projected dispersion of the dark matter itself cannot be observed. What is measured is that of a tracer population,
the galaxies of a group or cluster or the stars of a galaxy, embedded in the halo. For a tracer with density $\nu(r)$ in
a spherical mass distribution $M(r)$, the exchange of integration order of Sect.~\ref{sec:model} gives, for an isotropic
velocity distribution,
\begin{equation}\label{eq:tracer}
\Sigma_\nu\sigma_\los^2(R)=2\int_R^\infty\frac{\nu(r)\,M(r)}{r^2}\sqrt{r^2-R^2}\dd r ,
\end{equation}
with $\Sigma_\nu$ the surface density of the tracer, and the aperture dispersion follows from the corresponding moment
of $(r^2-R^2)^{3/2}$. For a Hernquist tracer, $\nu\propto1/[r(r+a)^3]$, or a Jaffe tracer, $\nu\propto1/[r^2(r+a)^2]$,
with scale radius $a$ (in units of $r_s$) in an NFW halo, the integrand of \eqref{eq:tracer} has the same structure
as that of $J_1$: a rational function times $\ln(1+r)$, plus a rational function, times the kernel. In the
logarithmic part only a pole moves, from $r=-1$ to $r=-a$, so Hermite's reduction leaves a simple term proportional
to $1/[(r+a)\sqrt{r^2-R^2}]$ in place of $1/[(1+r)\sqrt{r^2-R^2}]$; the rational part gives elementary functions, as
for $J_1$. The new logarithmic integral, $\int_R^\infty\ln(1+r)\dd r/[(r+a)\sqrt{r^2-R^2}]$, is again of weight two and can be treated with substitutions of the kind used in
Appendix~\ref{app:T1low}. Such two-component models are closer to what is observed than the single-component halo,
and in practice they are evaluated numerically \citep[e.g.][]{larkin2016darkmatter}.

For an anisotropic velocity distribution the line-of-sight projection is again a single quadrature of $\nu M$, against a kernel that is
elementary for the usual anisotropy profiles \citep[Eqs.~A15--A16]{mamon2005darkb}. The kernel brings its own
logarithms or inverse trigonometric functions, so functions of weight three, such as trilogarithms, can appear. The
same route should apply to cored haloes such as the Burkert \citep{burkert1995structure} and dPIE
\citep{eliasdottir2007where,bergamini2019enhanced} profiles, and to the two-component J3 and JJ models, whose
projections are stated not to be elementary \citep{ciotti2018twocomponent,ciotti2019new}. Preliminary calculations
support such extensions, both to other tracer populations and to other anisotropy prescriptions; we leave the details
to future work.

\section{Conclusions}\label{sec:conclusion}

The line-of-sight and aperture velocity dispersions of the isotropic, untruncated NFW halo, the most widely used model
for dark matter haloes, have closed forms in terms of elementary functions and two non-elementary functions. On
either side of $R=r_s$ the first is written with Clausen's function or its hyperbolic counterpart, in forms that
mirror the known surface density, and the second with the inverse tangent integral. The closed forms give
$\sigma_\los^2(r_s)=(49/2-3\pi/4-15\Cat-12\ln2)\,4\pi G\rho_sr_s^2$, the exact central and large-radius behaviour, the
maximum of $\sigma_\los$ at $R=0.510049\,r_s$ and the minimum $0.6154253$ of the aperture-to-virial ratio at
$c=3.23357$, and they serve as exact reference solutions for numerical projections. A quadrature that starts from the
Jeans equation and uses no step of the derivation agrees with them to about $10^{-31}$ over nine decades in radius. As
one application, the standard fit to the aperture-to-virial ratio is less accurate than stated for $1<c<5.7$. The
route to the closed forms, through projection identities, Hermite's reduction, rationalising substitutions and
polylogarithm identities, is systematic, and it should extend to stellar tracers in NFW haloes, whose projected
dispersions are what observations measure.

\section*{Data availability}
The verification scripts (Python with mpmath, SymPy, NumPy and SciPy) and their reference outputs are available as
ancillary files of the arXiv version of this paper and from the authors on request.

\begin{acknowledgements}
The computations used mpmath, SymPy, NumPy, SciPy and Matplotlib. A generative AI assistant was used, under A.A.'s direction, to derive and check the results, to write the verification scripts, and to draft the text. The authors are responsible for the content.
EvoMath is available for noncommercial use in \url{https://gitlab.com/evomath-group/evomath} and as a hosted service with a free tier in \url{https://evomath.ai}. A.A. is affiliated with EvoMath Research Labs UG, which develops EvoMath and provides the hosted service.

\end{acknowledgements}

\bibliographystyle{aa}
\bibliography{refs_aa}

\begin{appendix}
\nolinenumbers
\section{The functions $T_1$ and $T_2$}\label{app:T}

In the appendices we abbreviate $S=\sqrt{r^2-R^2}$. For $0<R<1$ we write $\lambda=\sqrt{1-R^2}$ and
$\tau=\arccosh(1/R)$; for $R>1$ we write $s=\sqrt{R^2-1}$, $\theta=\arccos(1/R)\in(0,\pi/2)$ and $\chi=\arccosh R$.
Thus $\tau$ and $\theta$ are the two branches of the variable $\vartheta$ of Sect.~\ref{sec:T}, and $F=\tau/\lambda$
for $R<1$ and $F=\theta/s$ for $R>1$.

\paragraph{The relations \eqref{eq:rel}.} For $z>0$, $\Li'(z)=-z^{-1}\ln(1-z)$ gives
$\frac{\dd}{\dd z}\big[\frac{\pi^2}{6}-\frac{z^2}{4}-\Li(e^{-z})\big]=-\frac z2-\ln(1-e^{-z})
=-\ln\big(2\sinh\frac z2\big)$. As $z\to0^+$ the bracket tends to $\pi^2/6-\Li(1)=0$, and the integral in
\eqref{eq:Clh} converges because $\ln(2\sinh\frac t2)=\ln t+O(t^2)$. This proves the relation for $\Clh$. The relation
for $\Cl$ is standard \citep[25.12.7 and 25.12.9]{dlmf}, and the one for $\Ti$ follows from
$\Li(iz)=-\int_0^zt^{-1}\ln(1-it)\dd t$ and $\operatorname{Im}\ln(1-it)=-\arctan t$.

\subsection{$T_1$ for $0<R<1$}\label{app:T1low}

\paragraph{Substitution.} Put $a=e^{-\tau}\in(0,1)$, so that $R=1/\cosh\tau=2a/(1+a^2)$ and
$\lambda=\tanh\tau=(1-a^2)/(1+a^2)$. The map $u\mapsto r=\frac R2(u+u^{-1})$ is increasing from $[1,\infty)$ onto
$[R,\infty)$, and
\begin{gather}
S=\frac R2\big(u-u^{-1}\big),\qquad 1+r=\frac{R\,(u+a)(u+a^{-1})}{2u},\notag\\
\frac{\dd r}{(1+r)\,S}=\frac1\lambda\Big[\frac{1}{u+a}-\frac{1}{u+a^{-1}}\Big]\dd u ,\label{eq:A1}
\end{gather}
since $\dd r/S=\dd u/u$ and $\frac2R(a^{-1}-a)^{-1}=\lambda^{-1}$. With $P=\ln(u+a)$, $Q=\ln(u+a^{-1})$, $L=\ln u$ and
$\ell=\ln(R/2)$ we have $\ln(1+r)=\ell+P+Q-L$, hence
\begin{equation}\label{eq:A2}
\lambda T_1=\int_1^\infty(\ell+P+Q-L)\,(P'-Q')\dd u .
\end{equation}

\paragraph{Antiderivative.} Let $k=a^{-1}-a>0$ and
\begin{equation}\label{eq:A3}
\begin{split}
\mathcal{A}(u)={}&P^2-Q^2+\Li\Big(\frac{-k}{u+a}\Big)-\Li\Big(\frac{k}{u+a^{-1}}\Big)\\
&+\ell\,(P-Q)+L\big[\ln(1+au)-\ln(1+u/a)\big]\\
&+\Li(-au)-\Li(-u/a).
\end{split}
\end{equation}
For $u\ge1$ all four dilogarithm arguments are real and below $1$, so $\mathcal{A}$ is real-analytic on $[1,\infty)$.
From $\Li'(z)=-z^{-1}\ln(1-z)$, $1+\frac{k}{u+a}=\frac{u+a^{-1}}{u+a}$ and $1-\frac{k}{u+a^{-1}}=\frac{u+a}{u+a^{-1}}$,
\begin{gather*}
\frac{\dd}{\dd u}\Li\Big(\frac{-k}{u+a}\Big)=(Q-P)P',\\ \frac{\dd}{\dd u}\Li\Big(\frac{k}{u+a^{-1}}\Big)=(P-Q)Q',\\
\frac{\dd}{\dd u}\big[\Li(-au)-\Li(-u/a)\big]=-\frac{D(u)}{u},
\end{gather*}
with $D(u)=\ln(1+au)-\ln(1+u/a)$.
Hence $\frac{\dd}{\dd u}[P^2-Q^2]+(Q-P)P'-(P-Q)Q'=(P+Q)(P'-Q')$, the $L$ terms give
$L\,[\frac{a}{1+au}-\frac{1}{a+u}]=-L(P'-Q')$, and $\mathcal{A}'=(\ell+P+Q-L)(P'-Q')$. Therefore
$\lambda T_1=\lim_{u\to\infty}\mathcal{A}(u)-\mathcal{A}(1)$.

\paragraph{The limit $u\to\infty$.} As $u\to\infty$, $P-Q=O(u^{-1})$ and $P+Q=O(\ln u)$, so $P^2-Q^2$ and $\ell(P-Q)$
vanish, and so do the two dilogarithms with arguments $\mp k/(u+a^{\pm1})$. By the inversion formula
$\Li(-v)+\Li(-1/v)=-\pi^2/6-\frac12\ln^2v$ for $v>0$ \citep[25.12.4]{dlmf},
$\Li(-au)-\Li(-u/a)=-2\ln a\ln u+O(u^{-1})$, while $L[\ln(1+au)-\ln(1+u/a)]=2\ln a\ln u+O(u^{-1}\ln u)$. The leading
terms cancel, so $\mathcal{A}(u)\to0$ and $\lambda T_1=-\mathcal{A}(1)$.

\paragraph{The value at $u=1$.} Using $k/(1+a)=(1-a)/a$, $k/(1+a^{-1})=1-a$, $P(1)=\ln(1+a)$, $Q(1)=\ln(1+a)-\ln a$
and $\ell=\ln a-\ln(1+a^2)$,
\begin{multline*}
\mathcal{A}(1)=2\ln a\ln(1+a)-\ln a\ln(1+a^2)\\+\Li(1-a^{-1})-\Li(1-a)+\Li(-a)-\Li(-a^{-1}).
\end{multline*}
We apply, in this order, the Landen-type formula $\Li(1-a)+\Li(1-a^{-1})=-\frac12\ln^2a$ \citep[25.12.3]{dlmf}, the
inversion formula $\Li(-a^{-1})=-\pi^2/6-\frac12\ln^2a-\Li(-a)$, Euler's reflection formula
$\Li(1-a)=\pi^2/6-\ln a\ln(1-a)-\Li(a)$ \citep[25.12.6]{dlmf} and the duplication formula $2\Li(a)+2\Li(-a)=\Li(a^2)$
\citep[25.12.5]{dlmf}:
\begin{align*}
\mathcal{A}(1)&=2\ln a\ln(1+a)-\ln a\ln(1+a^2)\\&\quad-2\Li(1-a)+2\Li(-a)+\frac{\pi^2}{6}\\
&=2\ln a\ln(1-a^2)-\ln a\ln(1+a^2)\\&\quad+\Li(a^2)-\frac{\pi^2}{6}.
\end{align*}
Hence
\begin{equation}\label{eq:A6}
\lambda T_1=\frac{\pi^2}{6}-\Li(a^2)+\ln a\,\ln\frac{1+a^2}{(1-a^2)^2}.
\end{equation}
Since $1+a^2=2a/R$ and $1-a^2=2a\lambda/R$, $\ln[(1+a^2)/(1-a^2)^2]=\tau-\ln(2\lambda^2/R)$. With $\ln a=-\tau$,
$a^2=e^{-2\tau}$ and \eqref{eq:rel} at $z=2\tau$, \eqref{eq:A6} becomes
\begin{equation}\label{eq:T1low}
\lambda\,T_1=\Clh(2\tau)+\tau\ln\frac{2\lambda^2}{R},
\end{equation}
which is \eqref{eq:T1} for $R<1$, because $F=\tau/\lambda$.

\subsection{$T_2$ for all $R$}\label{app:T2}

With $r=R\sec\psi$, $\psi\in[0,\pi/2)$, we have $S=R\tan\psi$, $\dd r/(rS)=\dd\psi/R$ and $1+r=(R+\cos\psi)/\cos\psi$,
so
\begin{equation}\label{eq:A7}
R\,T_2=\int_0^{\pi/2}\ln\Big(1+\frac{R}{\cos\psi}\Big)\dd\psi .
\end{equation}
Differentiating under the integral sign (on $R\ge R_0>0$ the derivative $1/(R+\cos\psi)$ of the integrand is bounded
by $1/R_0$), $\frac{\dd}{\dd R}[R\,T_2]=\int_0^{\pi/2}(R+\cos\psi)^{-1}\dd\psi$, and the substitution $r=R\sec\psi$ turns
this into $\int_R^\infty\dd r/[(1+r)S]=F(R)$ (Appendix~\ref{app:Ffh}). The integrand of \eqref{eq:A7} decreases to $0$
as $R\downarrow0$, so $R\,T_2\to0$ by monotone convergence, and $R\,T_2=\int_0^RF(R')\dd R'$. In particular $R\,T_2$ is
continuous on $[0,\infty)$.

\paragraph{$0<R<1$.} Let $g_-(R)=2\Ti(a)+\tau\arcsin R$ with $a=e^{-\tau}=R/(1+\lambda)$. Since
$\tan\frac\beta2=\sin\beta/(1+\cos\beta)$, $a=\tan(\frac12\arcsin R)$, that is $2\arctan a=\arcsin R$. With
$\Ti'(y)=y^{-1}\arctan y$, $\frac{\dd a}{\dd R}=\frac{a}{R\lambda}$ and $\frac{\dd\tau}{\dd R}=-\frac{1}{R\lambda}$,
\[
g_-'(R)=\frac{2\arctan a-\arcsin R}{R\lambda}+\frac{\tau}{\lambda}=\frac{\tau}{\lambda}=F(R).
\]
As $R\to0^+$, $a\to0$ and $\tau\arcsin R=O(R\ln\frac1R)$, so $g_-(0^+)=0$. Hence $g_-=R\,T_2$ on $(0,1)$, which is
the first line of \eqref{eq:T2}, and $T_2(1)=g_-(1^-)=2\Ti(1)=2\Cat$.

\paragraph{$R>1$.} Let $g_+(R)=2\Ti(q)+\frac\pi2\chi$ with $q=e^{-\chi}$. Since $\arctan q+\arctan q^{-1}=\pi/2$ and
$\arccos(1/\cosh\chi)=2\arctan e^{\chi}-\pi/2$ (both sides vanish at $\chi=0$ and have derivative $1/\cosh\chi$),
$2\arctan q=\frac\pi2-\theta$. With $\frac{\dd\chi}{\dd R}=\frac1s$ and $\frac{\dd q}{\dd R}=-\frac qs$,
\[
g_+'(R)=\frac{\pi/2-2\arctan q}{s}=\frac{\theta}{s}=F(R),
\]
and $g_+(1^+)=2\Ti(1)=2\Cat=T_2(1)$. Hence $g_+=R\,T_2$ on $(1,\infty)$, which is the second line of \eqref{eq:T2}.

\subsection{Analyticity, and $T_1$ for $R\ge1$}\label{app:Tcont}

\paragraph{Step 1: analyticity.} With $r=R\cosh t$,
\begin{gather*}
T_1(R)=\int_0^\infty\frac{\ln(1+R\cosh t)}{1+R\cosh t}\dd t,\\ T_2(R)=\int_0^\infty\frac{\ln(1+R\cosh t)}{R\cosh t}\dd t .
\end{gather*}
For $\operatorname{Re}R\ge\delta>0$ and $|R|\le K$, $|1+R\cosh t|\ge1+\delta\cosh t$ and $|\ln(1+R\cosh t)|\le\ln(1+K\cosh
t)+\pi$, so both integrands are bounded by an integrable function of $t$ that does not depend on $R$. Both integrands
are holomorphic in $R$, hence $T_1$ and $T_2$ are holomorphic on $\operatorname{Re}R>0$.

\paragraph{Step 2: the function $B$.} Let $B(a)=\Li(1-a^2)+\ln a\,\ln(1+a^2)$, holomorphic on $\operatorname{Re}a>0$.
For $0<a<1$ Euler's reflection formula gives $\pi^2/6-\Li(a^2)=\Li(1-a^2)+2\ln a\ln(1-a^2)$, so the right-hand side of
\eqref{eq:A6} equals $B(a)$ and $\lambda T_1=B(a)$. The Landen-type formula with $z=a^2$ gives
$\Li(1-a^{-2})=-\Li(1-a^2)-2\ln^2a$, hence $B(1/a)=-B(a)$ for $a>0$, and by the identity theorem for all
$\operatorname{Re}a>0$.

\paragraph{Step 3: $T_1$ as a function of $1-R^2$.} For $|\lambda|<1$ let $a(\lambda)=[(1-\lambda)/(1+\lambda)]^{1/2}$
(principal root). The map $\lambda\mapsto(1-\lambda)/(1+\lambda)$ sends the unit disc onto the right half-plane, so
$a$ is holomorphic on the disc with $\operatorname{Re}a>0$ and $a(-\lambda)=1/a(\lambda)$; for $\lambda=\tanh\tau$,
$a(\lambda)=e^{-\tau}$. By Step 2, $b(\lambda)=B(a(\lambda))$ is holomorphic and odd on the disc, so
$b(\lambda)/\lambda=\Phi(\lambda^2)$ with $\Phi$ holomorphic on the unit disc. By Appendix~\ref{app:T1low},
$T_1(R)=\Phi(1-R^2)$ for $0<R<1$, and since both sides are real-analytic on $0<R<\sqrt2$, also there.

\paragraph{Step 4: $R=1$.} $T_1(1)=\Phi(0)=b'(0)$. Since $a(\lambda)=1-\lambda+O(\lambda^2)$,
$\Li(1-a^2)=2\lambda+O(\lambda^2)$ and $\ln a\ln(1+a^2)=-\lambda\ln2+O(\lambda^2)$, so $T_1(1)=2-\ln2$.

\paragraph{Step 5: $1<R<\sqrt2$.} Take $\lambda=is$ with $s=\sqrt{R^2-1}\in(0,1)$. Then $(1-is)/(1+is)=e^{-2i\theta}$
with $\theta=\arctan s=\arccos(1/R)\in(0,\pi/4)$, so $a(is)=e^{-i\theta}$ and $T_1(R)=B(e^{-i\theta})/(is)$. Put
$z=e^{-2i\theta}$. Neither $z$ nor $1-z$ is real, so Euler's reflection formula holds at $z$ by analytic continuation.
With principal logarithms, $\ln z=-2i\theta$, $\ln a=-i\theta$,
\begin{gather*}
\ln(1-z)=\ln(2\sin\theta)+i\Big(\frac\pi2-\theta\Big),\\ \ln(1+z)=\ln(2\cos\theta)-i\theta,
\end{gather*}
and the Fourier series \citep[25.12.7--8]{dlmf} give $\Li(z)=\pi^2/6-\theta(\pi-\theta)-i\Cl(2\theta)$. Hence
\begin{align*}
\Li(1-z)&=\frac{\pi^2}{6}-\ln z\ln(1-z)-\Li(z)\\&=\theta^2+2i\theta\ln(2\sin\theta)+i\Cl(2\theta),\\
\ln a\,\ln(1+a^2)&=-i\theta\ln(2\cos\theta)-\theta^2,\\
B(e^{-i\theta})&=i\Big[\Cl(2\theta)+\theta\ln\frac{2\sin^2\theta}{\cos\theta}\Big].
\end{align*}
With $\cos\theta=1/R$ and $\sin\theta=s/R$, $2\sin^2\theta/\cos\theta=2s^2/R$, which gives
\begin{equation}\label{eq:T1high}
s\,T_1=\Cl(2\theta)+\theta\ln\frac{2s^2}{R},
\end{equation}
which is \eqref{eq:T1} for $R>1$, because $F=\theta/s$. So far this holds for $1<R<\sqrt2$.

\paragraph{Step 6: all $R>1$.} The right-hand side of \eqref{eq:T1high} is real-analytic on $(1,\infty)$, because
$2\theta\in(0,\pi)$ and $\Cl$ is real-analytic on $(0,2\pi)$. $T_1$ is real-analytic by Step 1. They agree on
$(1,\sqrt2)$, hence on $(1,\infty)$.

\section{Reduction of $J_1$ and $J_3$}\label{app:J}

Write $w_k=S^k/[r^3(1+r)^2]$, so that by \eqref{eq:Jk}
$J_k=\int_R^\infty\ln(1+r)\,w_k\dd r-\int_R^\infty r\,w_k(1+r)^{-1}\dd r$. The
reduction uses three rational-function identities in $r$, whose data are listed in Table~\ref{tab:hermite}. Since
$\frac{\dd}{\dd r}[U\,S]=[(r^2-R^2)U'+rU]/S$ for a rational function $U(r)$, each identity becomes a rational identity
after multiplication by $S$, and can be checked by differentiation.

\paragraph{Step 1 (Hermite decomposition).} With $H_k=\Pi_k(r)\,S/[r^2(1+r)]$,
\begin{equation}\label{eq:hermite}
w_k=H_k'+\frac{c_1^{(k)}}{(1+r)S}+\frac{c_2^{(k)}}{rS}.
\end{equation}
Multiplying by $\ln(1+r)$ and integrating, the last two terms give $c_1^{(k)}T_1+c_2^{(k)}T_2$.

\paragraph{Step 2 (integration by parts).} $H_k$ vanishes at $r=R$ because $S$ does, and $H_k=C_k+O(r^{-1})$ for
$r\to\infty$. Hence
\begin{gather*}
J_k=c_1^{(k)}T_1+c_2^{(k)}T_2+C_k\ln(1+R)\\-\int_R^\infty\Big[S\,Q_k-\frac{C_k}{1+r}\Big]\dd r,\\
Q_k=\frac{\Pi_k}{r^2(1+r)^2}+\frac{S^{k-1}}{r^2(1+r)^3},
\end{gather*}
where $Q_k$ is rational because $S^{k-1}=(r^2-R^2)^{(k-1)/2}$.

\paragraph{Step 3 (algebraic part).} With the rational $V_k$ of Table~\ref{tab:hermite},
\begin{equation}\label{eq:alg}
S\,Q_k=\frac{\dd}{\dd r}\big[S\,V_k\big]+\frac{C_k}{S}+\frac{\beta_k}{rS}+\frac{\gamma_k}{(1+r)S}.
\end{equation}
The coefficient of $1/S$ equals $C_k$, which makes the integral of Step 2 converge. $S\,V_k$ vanishes at $r=R$ and tends
to $L_k$ as $r\to\infty$.

\paragraph{Step 4 (elementary integrals).} For $R>0$, $\int_R^\infty[S^{-1}-(1+r)^{-1}]\dd r=\ln[2(1+R)/R]$ (antiderivative
$\arccosh(r/R)-\ln(1+r)$), $\int_R^\infty\dd r/(rS)=\pi/(2R)$ (with $r=R\sec\psi$) and $\int_R^\infty\dd r/[(1+r)S]=F(R)$.
Assembling Steps 1 to 4,
\begin{equation}\label{eq:assembly}
J_k=c_1^{(k)}T_1+c_2^{(k)}T_2+C_k\ln\frac R2-L_k-\frac{\pi\beta_k}{2R}-\gamma_kF .
\end{equation}
With the data of Table~\ref{tab:hermite} and $F=1-\frac12(R^2-1)\,\Sigma$, this gives \eqref{eq:J1} and \eqref{eq:J3}.
For $J_1$ the constant parts combine as $(7R^2-6+R^4+3R^2-5)/[2(R^2-1)]=(11+R^2)/2$; for $J_3$ they give
$(3-11R^2-R^4)/2$, and the coefficient of $\Sigma$ is $(R^2-1)(R^4+4R^2-2)/4$. All identities are rational in $r$ and
$R$ and the integrals of Step 4 hold for every $R>0$, so the reduction is the same on both sides of $R=1$. At $R=1$
the formulas hold by continuity: $J_k$ is continuous by dominated convergence, and the right-hand sides are
real-analytic.

\begin{table*}
\caption{Data of the rational identities \eqref{eq:hermite} and \eqref{eq:alg}. $C_k=\lim_{r\to\infty}H_k$ and
$L_k=\lim_{r\to\infty}S\,V_k$.}
\label{tab:hermite}
\centering
\renewcommand{\arraystretch}{1.45}
\begin{tabular}{lll}
\hline\hline
 & $k=1$ & $k=3$\\
\hline
$\Pi_k(r)$ & $\tfrac12(6r^2+3r-1)$ & $\tfrac12\big[(2-6R^2)r^2-3R^2r+R^2\big]$\\
$c_1^{(k)}$ & $3R^2-2$ & $3R^2(1-R^2)$\\
$c_2^{(k)}$ & $-(3R^2-\tfrac12)$ & $\tfrac32R^2(2R^2-1)$\\
$C_k$ & $3$ & $1-3R^2$\\
$V_k(r)$ & $\dfrac{(6-7R^2)r^2+(8-9R^2)r+1-R^2}{2r(R^2-1)(1+r)^2}$ & $\dfrac{(7R^2-1)r^2+(9R^2-2)r+R^2}{2r(1+r)^2}$\\
$\beta_k$ & $R^2/2$ & $-R^4/2$\\
$\gamma_k$ & $\dfrac{5-3R^2-R^4}{2(R^2-1)}$ & $\tfrac12(R^4+4R^2-2)$\\
$L_k$ & $\dfrac{6-7R^2}{2(R^2-1)}$ & $\tfrac12(7R^2-1)$\\
\hline
\end{tabular}
\end{table*}

\section{The functions $F$, $\Sigma$ and $M_p$}\label{app:Ffh}

With $r=R\cosh t$, $\int_R^\infty\dd r/[(1+r)S]=\int_0^\infty(1+R\cosh t)^{-1}\dd t$, which is elementary and equals the
function $F$ of \eqref{eq:F}. For $R<1$, $F=\tau/\lambda$ with $\tanh\tau=\lambda$, so for $|1-R^2|<1$ and
$\lambda^2=1-R^2$,
\[
F=\frac{\artanh\lambda}{\lambda}=\sum_{n\ge0}\frac{\lambda^{2n}}{2n+1},\qquad
\frac{1-F}{R^2-1}=\sum_{n\ge0}\frac{\lambda^{2n}}{2n+3},
\]
and the continuation $\lambda=is$ gives the form for $R>1$. Differentiating $\tau/\lambda$ gives
$F'=(1-R^2F)/[R(R^2-1)]$. For $b>0$ the scaling $r=by$ gives $\int_R^\infty\dd r/[(b+r)S]=b^{-1}F(R/b)$; differentiating
in $b$ at $b=1$ gives $\int_R^\infty\dd r/[(1+r)^2S]=F+RF'=(1-F)/(R^2-1)$. Since $\rho\,r=(1+r)^{-2}$, this is
$\Sigma/2$, which proves \eqref{eq:Sigma} and gives the series
\[
\Sigma(R)=2\sum_{n\ge0}\frac{(1-R^2)^n}{2n+3}\qquad(|1-R^2|<1).
\]
Next, $\frac{\dd}{\dd R}[\ln(R/2)+F]=R\,\Sigma/2$ and $\ln(R/2)+F\to0$ as $R\to0$, so $M_p=4\pi[\ln(R/2)+F]$, which is
\eqref{eq:Mp}. Hence $F$, $\Sigma$ and $M_p$ are real-analytic on $(0,\infty)$, with $F(1)=1$, $\Sigma(1)=\frac23$ and
$M_p(1)=4\pi(1-\ln2)$. For $R\to0$, $\Sigma=2[\ln(2/R)-1]+O(R^2\ln\frac1R)$ and
$M_p=2\pi R^2[\ln(2/R)-\frac12]+O(R^4\ln\frac1R)$; for $R\to\infty$, $F=\pi/(2R)+O(R^{-2})$.

\section{Values and limits}\label{app:limits}

\paragraph{$R=1$.} Insert $T_1(1)=2-\ln2$, $T_2(1)=2\Cat$ and $\Sigma(1)=\frac23$ into \eqref{eq:J1} and \eqref{eq:J3}.
For $J_1$ the coefficients of $T_1$ and $T_2$ are $1$ and $-\frac52$ and the elementary part is
$6-\pi/4-3\ln2+\frac16$; for $J_3$ they are $0$ and $\frac32$ and the elementary part is $-\frac92+\pi/4+2\ln2$. Hence
\begin{align}
J_1(1)&=\frac{49}{6}-\frac\pi4-5\,\Cat-4\ln2=0.0288518101433,\notag\\
J_3(1)&=-\frac92+\frac\pi4+3\,\Cat+2\ln2=0.419589307049,\label{eq:J1at1}
\end{align}
and \eqref{eq:obs} with $\Sigma(1)=\frac23$ and $M_p(1)=4\pi(1-\ln2)$ gives \eqref{eq:rs}.

\paragraph{Centre.} $J_1(0^+)=(4\pi)^{-1}\int_0^\infty M(r)\,r^{-2}(1+r)^{-2}\dd r$ by monotone convergence
($S\uparrow r$ as $R\downarrow0$). The function
\begin{multline*}
\mathcal{B}(r)=\ln^2(1+r)+2\Li(-r)-\frac{\ln(1+r)+2}{1+r}\\-\frac{1}{2(1+r)^2}-\frac{\ln(1+r)}{r}
\end{multline*}
satisfies $\mathcal{B}'=M/[4\pi r^2(1+r)^2]$, $\mathcal{B}(0^+)=-\frac72$, and by the inversion formula
$2\Li(-r)+\ln^2(1+r)\to-\pi^2/3$ as $r\to\infty$ while the other terms vanish. Hence $J_1(0^+)=\frac72-\frac{\pi^2}{3}$.
In the same way $J_3(0^+)=(4\pi)^{-1}\int_0^\infty\rho\,M\,r\dd r=\frac12$.

\paragraph{Order $R$.} Write
$4\pi[J_1(R)-J_1(0)]=\int_R^\infty M\,(S-r)\,r^{-3}(1+r)^{-2}\dd r-\int_0^RM\,r^{-2}(1+r)^{-2}\dd r$.
Since $M(r)=2\pi r^2+O(r^3)$, the second integral is $2\pi R+O(R^2)$. With $r=Ry$ the first becomes $4\pi R$ times an
integral over $y\ge1$ whose integrand is bounded by $\frac12(1-\sqrt{1-y^{-2}})$ (because $0\le M(r)\le2\pi r^2$) and
tends pointwise to $\frac12(\sqrt{1-y^{-2}}-1)$. By dominated convergence the first integral is
$-2\pi R\int_1^\infty(1-\sqrt{1-y^{-2}})\dd y+o(R)=-2\pi R\,(\frac\pi2-1)+o(R)$. Together, $J_1=J_1(0)-\pi R/4+o(R)$.

\paragraph{Order $R^2$.} Put $\ell_0=\ln(2/R)$. With $\lambda=1-R^2/2+O(R^4)$, $\tau=\ell_0-R^2/4+O(R^4)$,
$\ln(2\lambda^2/R)=\ell_0-R^2+O(R^4)$ and $\Li(a^2)=R^2/4+O(R^4)$, \eqref{eq:A6} gives
$\lambda T_1=\pi^2/6-R^2/4-\frac34\ell_0R^2+O(R^4\ell_0)$. For $T_2$, the first line of \eqref{eq:T2} with
$a=R/2+R^3/8+O(R^5)$, $\Ti(a)=a-a^3/9+O(a^5)$ and $\arcsin R=R+R^3/6+O(R^5)$ gives
$R\,T_2=R(1+\ell_0)+R^3(\ell_0/6-1/36)+O(R^5\ell_0)$. With $F=\tau/\lambda$ and $\Sigma=2(F-1)/(1-R^2)$,
\begin{gather*}
T_1=\frac{\pi^2}{6}+R^2\Big[\frac{\pi^2}{12}-\frac14-\frac34\ell_0\Big],\\
T_2=1+\ell_0+R^2\Big[\frac{\ell_0}{6}-\frac{1}{36}\Big],\\
\frac\Sigma2=\ell_0-1+R^2\Big[\frac32\ell_0-\frac54\Big],
\end{gather*}
each up to $O(R^4\ell_0^2)$. Substituting in \eqref{eq:J1}, the constant terms give $\frac72-\frac{\pi^2}{3}$, the
coefficient of $R^2\ell_0$ is $\frac32-3+\frac1{12}-\frac32+\frac{15}{4}=\frac56$, and the remaining $R^2$ terms give
$\frac{\pi^2}{3}+\frac12-3-\frac1{72}+\frac12+\frac32-\frac{25}{8}=\frac{\pi^2}{3}-\frac{131}{36}$. All ingredients other
than the term $-\pi R/4$ are even in $R$ at fixed $\ell_0$, so no $R^3$ term occurs, and
\begin{multline}\label{eq:smallJ}
J_1(R)=\frac72-\frac{\pi^2}{3}-\frac{\pi R}{4}+R^2\Big[\frac56\ln\frac2R+\frac{\pi^2}{3}-\frac{131}{36}\Big]\\
+O\Big(R^4\ln^2\frac1R\Big).
\end{multline}
The approach to the $R^2$ term is slow: the ratio $[J_1-J_1(0)+\pi R/4]/(R^2\ln(1/R))$ equals $0.883$, $0.866$,
$0.858$ and $0.850$ at $R=10^{-2}$, $10^{-3}$, $10^{-4}$ and $10^{-6}$, against the limit $5/6$.

\paragraph{Dispersions at small $R$.} By \eqref{eq:smallJ},
$\frac13[\frac12-J_3]=\int_0^RR'J_1\dd R'=\frac{R^2}{2}J_1(0)-\frac{\pi R^3}{12}+O(R^4\ln\frac1R)$. Dividing $J_1$ by
$\Sigma/2=\ln(2/R)-1+O(R^2\ln\frac1R)$, and $\frac13[\frac12-J_3]$ by
$M_p/(4\pi)=\frac{R^2}{2}[\ln(2/R)-\frac12]+O(R^4\ln\frac1R)$ (Appendix~\ref{app:Ffh}), gives \eqref{eq:small} with
relative errors of order $R^2\ln(1/R)$.

\paragraph{Large $R$.} For $r\ge1$ write $M(r)/(4\pi)=\ln r-1+\varepsilon(r)$ and $r^2/(1+r)^2=1+\varrho(r)$ with
$\varepsilon=\ln(1+r^{-1})+(1+r)^{-1}\in(0,2/r)$ and $\varrho=-(2r+1)/(1+r)^2\in(-2/r,0)$. Then
$|M(1+\varrho)/(4\pi)-(\ln r-1)|\le C_0(1+\ln r)/r$ with an absolute constant $C_0$, and with $S\le r$,
\[
J_1=R^{-3}\int_1^\infty\frac{(\ln R+\ln y-1)\sqrt{y^2-1}}{y^5}\dd y+O\Big(\frac{\ln R}{R^4}\Big).
\]
With $y=\sec\psi$, $\int_1^\infty\sqrt{y^2-1}\,y^{-5}\dd y=\pi/16$ and
$\int_1^\infty\ln y\sqrt{y^2-1}\,y^{-5}\dd y=\frac{\pi\ln2}{16}-\frac{\pi}{64}$. For $J_3$ the same estimate with
$(r^2-R^2)^{3/2}\le r^3$ gives an error $O(R^{-2}\ln R)$, with
$\int_1^\infty(y^2-1)^{3/2}y^{-5}\dd y=3\pi/16$ and $\int_1^\infty\ln y\,(y^2-1)^{3/2}y^{-5}\dd y=\frac{3\pi}{16}(\ln2+\frac34)$.
Hence
\begin{align}
J_1(R)&=\frac{\pi}{16R^3}\Big[\ln(2R)-\frac54\Big]+O\Big(\frac{\ln R}{R^4}\Big),\notag\\
J_3(R)&=\frac{3\pi}{16R}\Big[\ln(2R)-\frac14\Big]+O\Big(\frac{\ln R}{R^2}\Big).\label{eq:largeJ}
\end{align}
Finally $\Sigma=2R^{-2}[1+O(R^{-1})]$ and $M_p=4\pi[\ln(R/2)+O(R^{-1})]$, which with \eqref{eq:obs} give
\eqref{eq:large}.

\section{Verification scripts}\label{app:scripts}

The scripts use Python~3 with mpmath~1.3, SymPy~1.14, NumPy, SciPy and Matplotlib, and run with
\texttt{OPENBLAS\_NUM\_THREADS=1}. Each writes a plain-text output file, and reference outputs are included. The
scripts \texttt{v01} to \texttt{v13} keep the symbols of an earlier version of this paper: $X$ and $x$ for $R$ and $r$,
$m=M/4\pi$, $f=\Sigma/2$, $h=M_p/4\pi$, and $\mathcal{S}(t)=\Clh(2t)$.
\texttt{nfw\_closed.py} is a reference implementation of the closed forms in mpmath at any precision.
\texttt{v01} checks the substitution \eqref{eq:A1}, the antiderivative \eqref{eq:A3} and \eqref{eq:A6} exactly, and
Sect.~\ref{sec:T} against quadrature at 50 digits. \texttt{v03} verifies every identity of Table~\ref{tab:hermite} and
the assembly \eqref{eq:assembly} exactly, and compares \eqref{eq:J1} and \eqref{eq:J3} with 40-digit quadrature.
\texttt{v04} gives the values at $R=1$, the limits, $\eta(c)$ and its minimum, and the maximum of $\sigma_\los$.
\texttt{v08} is the definition-based route of Sect.~\ref{sec:num} (about 80 minutes in total). \texttt{v09} checks the
steps of Appendices~\ref{app:T} to~\ref{app:limits} line by line. \texttt{v10} produces Tables~\ref{tab:values}
and~\ref{tab:mm07} and the concentrations where the error of the MM07 fit crosses $0.12\%$ and $5\%$. \texttt{v11}
gives the double-precision errors of Sect.~\ref{sec:num}, \texttt{v12} the $R^2$ term of \eqref{eq:smallJ}, and
\texttt{v13} the relation \eqref{eq:rel} for $\Clh$, the two branches of \eqref{eq:T2} and the derivative of $R\,T_2$ at
50 digits. \texttt{v14} codes the formulas of Sects.~\ref{sec:model} to~\ref{sec:app} as printed, in the notation of
this paper, and compares them with quadratures of their definitions and with \texttt{nfw\_closed.py} at 40 digits. The
figure scripts write the data behind Figs.~\ref{fig:validation} and~\ref{fig:profiles}. Apart from \texttt{v08}, each
script runs in under a minute.

\end{appendix}

\end{document}